\documentclass[aps,prl,twocolumn,showpacs,superscriptaddress,groupedaddress]{revtex4}  
\usepackage{graphicx}  
\usepackage{dcolumn}   
\usepackage{bm}
\usepackage{xcolor}        
\usepackage{amssymb}   
\usepackage{amsmath} 
\usepackage{verbatim}
\def\bea{\begin{eqnarray}}
\def\eea{\end{eqnarray}}

\begin{document}
\title{Read-Green points and level crossings in XXZ central spin models and $p_x+ip_y$ topological superconductors}
\author{Alexandre Faribault, Houda Koussir, Mohamed Houssein Mohamed}
\affiliation{Universit\'e de Lorraine, CNRS, LPCT, F-54000 Nancy, France}
\begin{abstract}

In this work, we study the full set of eigenstates of a $p_x+ip_y$ topological superconductor coupled to a particle bath which can be described in terms of an integrable Hamiltonian of the Richardson-Gaudin class. The results derived in this work also characterise the behaviour of an anisotropic XXZ central spin model in a external magnetic field since both types of Hamiltonian are know to share the exact same conserved quantities making them formally equivalent. 

We show how by ramping the coupling strength (or equivalently the magnetic field acting in the z-direction on the central spin), each individual eigenstate undergoes a sequence of gain/loss of excitations when crossing the specific values known as Read-Green points. These features are shown to be completely predictable, for every one of the $2^N$ eigenstates, using only two integers obtainable easily from the zero-coupling configuration which defines the eigenstate in question.

These results provide a complete map of the particle-number sectors (superconductor) or magnetisation sectors (central spin) involved in the large number of level-crossings which occur in these systems at the Read-Green points. It further allows us to define quenching protocols which could create states with remarkably large excitation-number fluctuations.

\end{abstract}

\pacs{}
\maketitle

\section{introduction}

Since its first explanation by Bardeen, Cooper and Schrieffer in 1957 \cite{bcs}, the theoretical description of superconducting systems has been vastly enriched by going beyond their original mean-field treatment of s-wave pairing interactions. An exact solution to the reduced s-wave BCS pairing hamiltonian was found by Richardson in 1963 \cite{richardson63,richardson64}, a result which saw an important surge in interest in the early 2000's \cite{sierra,vondelft} in the theoretical description of experiments on superconducting nanograins \cite{ralph1,ralph2,ralph3} .

It was also around that time that the s-wave pairing model was explicitly shown, in 1997, to be integrable  when Cambiaggio et al. \cite{cambiaggio_integrability_1997} explicitly found the set of commuting conserved operators defining its quantum integrability. Using an Anderson's pseudo-spin representation, this set of commuting operators then make the s-wave pairing problem equivalent to an isotropic XXX Gaudin magnet \cite{gaudin76,gaudinbook,ortiz}.

These ideas have then been built upon to build integrable pairing hamiltonians from anisotropic Richardson-Gaudin models \cite{amico,rombouts,lerma}. Integrable BCS pairing models with $p_x+i p_y$ symmetry have then been studied beyond the common mean-field approximation using the massive simplifications that integrability and the Bethe Ansatz solution can provide \cite{ibanez,dunning,raemdonck,links2015,shen}. 
Such models have a strong interest since they can exhibit topological superconductivity \cite{read,ryu,sato} whose occurence could possibly be exploited in quantum computational applications \cite{tewari,sau,dassarma}.

A recent result, upon which this work builds, is the observation by Claeys et al. \cite{claeys_readgreen_2016}. By coupling weakly such a $p+ip$ superconducting system to an external bath of particles we break the U(1)-symmetry which enforces the conservation of the number of Cooper pairs. In doing so, the ground state of the system, when raising the coupling constant $g$, will undergo a series of steps by systematically gaining a single Cooper pair when the coupling goes through specific values $g=g_i$ dubbed Read-Green points. The resulting ground state at, and around, these points then becomes a coherent superposition of an $M$ and $M+1$ Cooper pair states which exhibits pair number fluctuations. This is made possible by the weak coupling to the bath, which turns into avoided crossings, the level crossings between number conserving sectors which would occur at these specific couplings in a closed (number-conserving) system.

In this work, a similar study is carried out for every eigenstate of the system in order to characterise the behaviour of the full set of eigenstates across those Read-Green points. We first show explicitly that a step-like structure occurs over the (almost) complete Hilbert space and that it can be richer for the excited states than the one the ground state undergoes. Indeed, excited states can show both gains or losses of excitations when $g$ goes across a Read-Green point and these gains and losses can involve much more than a single excitation. Secondly, we demonstrate that the sequence of gains and losses  can be completely predicted using only two, state-specific, integers which are then sufficient to know the complete profile of excitation-number that each individual eigenstate goes through as the coupling is varied. Finally, through this full understanding of the involved (avoided) crossings, we discuss a quenching protocol designed to create specific states which should allow remarkably large number fluctuations by hybridising two sectors at filling factors $\rho \approx 0$ and $\rho \approx 1$.




\section{Richardson-Gaudin models}

The integrability of the $p+ip$ pairing models is fundamentally linked to the fact that they can be built as a linear combination of the set of $N$ mutually commuting operators given, in Anderson pseudo-spin representation, by:
\bea
\tilde{R}_i  = \frac{1}{g}\sigma^z_i  + \sum_{j \ne i}^N \left[X_{ij} \left(\sigma^x_i\sigma^x_j + \sigma^y_i\sigma^y_j\right) +Z_{ij} \sigma^z_i\sigma^z_j \right].
\label{charges}
\eea
Here $i=1,2, ... N$ labels one of the possible momenta $k_i$ at which one can either find a Cooper pair or not. In order to insure the commutation rules $[R_i,R_j]=0$ and consequently integrability, one needs to have $X_{ij}$ and $Z_{ij}$ parametrised as $X_{ij} = \frac{\sqrt{(\alpha \epsilon_i+\beta)(\alpha \epsilon_j+\beta)}}{\epsilon_i-\epsilon_j}$ and  $Z_{ij} = \frac{\alpha \epsilon_j+\beta}{\epsilon_i-\epsilon_j}$, for arbitrary parameters $\alpha$, $\beta$ and $(\epsilon_1 \dots \epsilon_N)$ \cite{ortiz,claeys_dimo,skrypnyk_review}.

Each of these individual conserved charges defines an anisotropic (XXZ) central spin model in which $i$ now labels each of the $N$ spins present. The operator $\tilde{R}_i$ then corresponds to an Hamiltonian in which the central spin, of index $i$, feels a $z$-oriented magnetic field (chosen here as $B_z = \frac{1}{g}$) and is also anisotropically coupled to each of the other $N-1$ individual spins. The fermionic $p+ip$ pairing hamiltonian is obtained through a well-documented \cite{amico,claeys_readgreen_2016,lukyanenko,ibanez,dunning,raemdonck,links2015,shen,skrypnyk_bcs} sum over these conserved charges using the Cooper-pair realisation of the SU(2) algebra which makes $\sigma^z_i = c^\dag_{k_i} c_{k_i}+c^\dag_{-k_i} c_{-k_i} -1$ while $\sigma^\pm_i$ creates or annihilates a Cooper pair in the $(k_i,-k_i)$ momentum state. The parameter $g$, which defines an external magnetic field in the central spin models, now plays the role of the pairing strength. Since both the pairing and the central spin model are defined by the same set of commuting conserved operators, they share the same eigenbasis which allows us to discuss the properties of the eigenstates of both models in the exact same terms. Throughout this work we will therefore use the term ``number of excitations" in order to describe either the total number of Cooper pairs in a pairing model or the total number of up-pointing spins in the central spin model.

The common eigenstates of the conserved charges (\ref{charges}), and therefore of the corresponding superconducting pairing model, are all such that they have a fixed total number of excitations since each of the $\tilde{R}_i$ operators also commute with the operator $\hat{M} = \frac{1}{2} \sum_{i=1}^N \sigma^z_i + 1$ whose eigenvalues $0,1,2 \dots N$ define this total number. This conservation reflects an underlying U(1)-symmetry. Adding an XY-plane component to the magnetic field or equivalently for the superconductor, by coupling it to an external particle bath will break this symmetry. Remarkably one can do so without breaking the integrability of the system \cite{lukyanenko,claeys_readgreen_2016,links2017,shen,skrypnyk_review} as, indeed, the following set of commuting operators:
\bea
&& R_i =  \frac{\gamma }{\sqrt{\alpha \epsilon_i+ \beta}} \sigma^x_i + \frac{\lambda }{\sqrt{\alpha \epsilon_i + \beta}} \sigma^y_i + \frac{1}{g}\sigma^z_i \nonumber\\&&+ \sum_{j \ne i}^N \left[X_{ij} \left(\sigma^x_i\sigma^x_j +\sigma^y_i\sigma^y_j \right)+Z_{ij} \sigma^z_i\sigma^z_j\right],
\label{conserved}
\eea
\noindent still commute with one another therefore defining an integrable model allowing us to retain the major simplifications that integrability has to offer. Our numerically study of the model's eigenstates makes use of recent work \cite{dimo,claeys_dimo,skrypnyk_review} which has shown explicitly that the set of eigenvalues $(r_1,r_2 \dots r_N)$ (of the operators $(R_1,R_2 \dots R_N)$ given in eq. (\ref{conserved})) which define each individual eigenstate, are also given by the set of solutions of a simple system of $N$ quadratic equations:
\bea
r_i^2 = \sum_{j \ne i} \Gamma_{ij} r_j +K_i,
\label{quadeq}
\eea
\noindent with $\Gamma_{ij} =2 \frac{\alpha \epsilon_j+ \beta}{\epsilon_i-\epsilon_j}$ and $K_i =   \frac{\gamma^2 + \lambda^2 }{\alpha \epsilon_i+ \beta}+ \frac{1}{g^2} + \displaystyle \sum_{j\ne i }^N \left( \frac{2(\alpha \epsilon_i+ \beta)(\alpha \epsilon_j+ \beta)+(\alpha \epsilon_j+ \beta)^2}{(\epsilon_i-\epsilon_j)^2}\right)$ \cite{dimo}. The knowledge of the eigenvalues $(r_1^n \dots r_N^n)$ associated to the eigenstate of index $n$: $\left|\psi_n\right>$, in conjunction with the quadratic equation they obey, gives a simple numerical access to the expectation values $\left<\psi_n\right|\sigma^\alpha_i\left|\psi_n\right>$ of every local spin operator $i=1 \dots N$, in any direction $\alpha=x,y,z$, by direct use of the Hellmann-Feynman theorem \cite{claeys_dimo}.



At any given value of $g$, i.e. of the magnetic field or the coupling strength, each individual eigenstate can be uniquely indexed by specifying its $g=0$ parent state. Indeed, each eigenstate at finite $g$ can be built by deforming a given $g=0$ eigenstate (parent state) by incrementing the coupling strength in small steps. The previously found eigenvalues  $(r_1(g-\Delta g) \dots r_N(g-\Delta g))$ provides an approximative solution for the eigenvalues at $g$ which, for $\Delta g$ small enough, stays within a particular solution's bassin of attraction of an iterative Newton-Raphson algorithm. By labelling the spins in such a way that $\epsilon_1 < \epsilon_2  < \epsilon_3 < \dots < \epsilon_N$, we will use the notation $\bullet$ for an up spin and $\circ$ for a down spin so that, for example, the parent ($g=0$) eigenstate $\left|\uparrow_1\right> \otimes \left|\downarrow_2\right> \otimes \left|\downarrow_3\right>  \otimes \left|\uparrow_4\right> \otimes \left|\downarrow_5\right> \ \dots \ \otimes \left|\uparrow_N\right>$ will be represented as $\bullet \circ \circ \bullet \circ \ \dots \ \bullet$, with the symbols ordered from left to right in increasing $\epsilon_i$ order. In the central spin model described by hamiltonian $R_1$, it means that $\sigma_1$ is considered the central spin while the environmental spins will be numbered in decreasing order of the magnitude of their coupling to the central spin, i.e.: the closer a spin is to the central one, the larger its coupling and therefore the lower its index. 

\section{Results}

\begin{figure}[h]
\includegraphics[width=7.5cm]{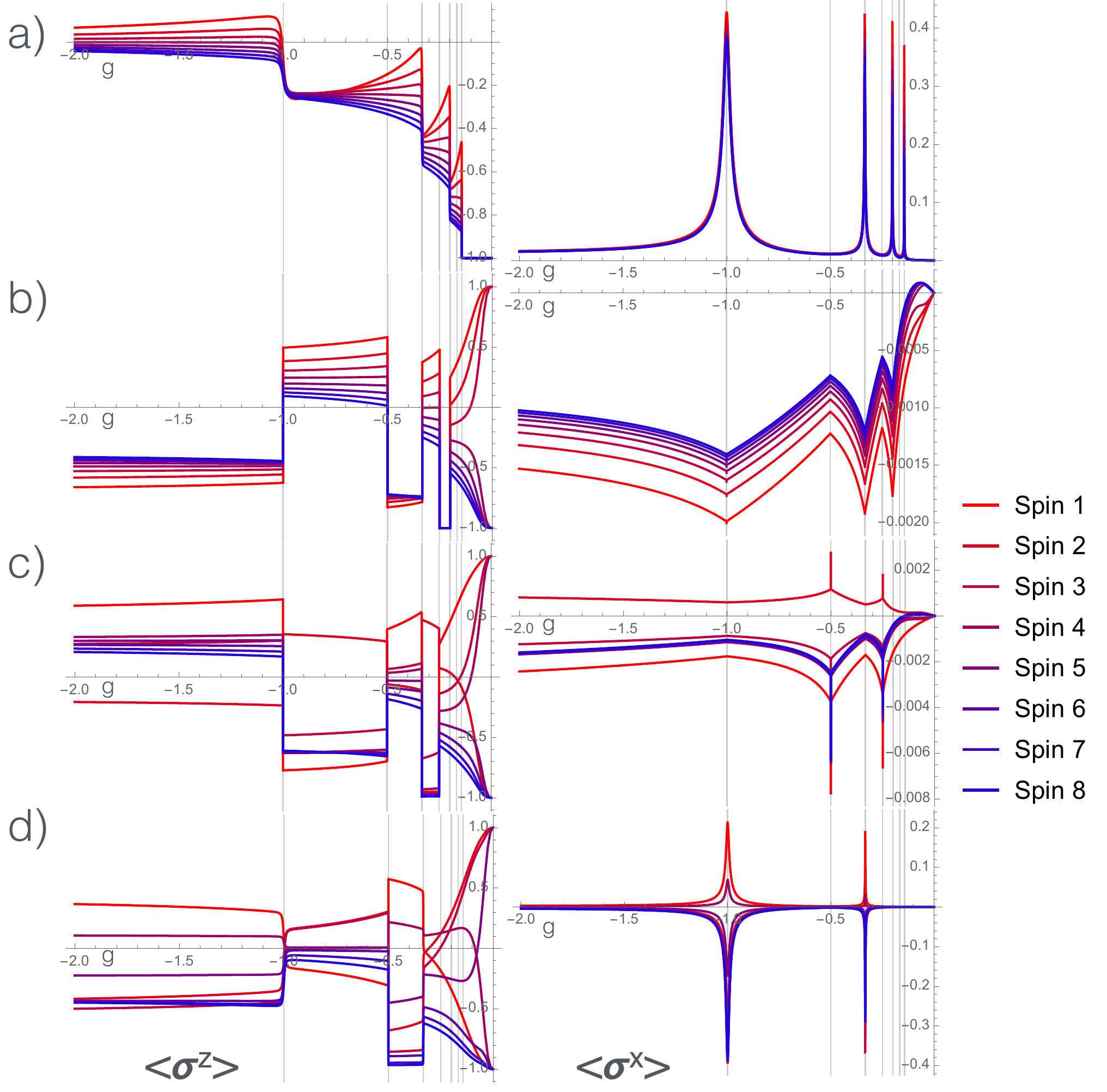} 
\vspace{-0.2cm}
\caption{Local expectation values $\left<\sigma_i^z\right>$ and $\left<\sigma_i^x\right>=\left<\sigma_i^y\right>$ for $\epsilon_i = i$, $\gamma= \lambda =0.005$ $\alpha = \beta =1$ for a selection of eigenstates, from top to bottom: a) $\circ \circ \circ \circ \circ \circ \circ \ \circ $, b) $ \bullet\bullet\bullet\circ\circ\circ\circ \ \circ $, c) $  \bullet\circ\bullet\bullet\circ\circ\circ \ \circ $, d) $ \circ \bullet\bullet\circ\bullet\circ\circ\ \circ $. The vertical lines mark the Read-Green points at $\frac{1}{|g|} = 7,6,5,4,3,2,1$. }
\label{sxsz}
\end{figure}

\begin{figure}[t]
\begin{center}
\includegraphics[height=17cm,angle=90]{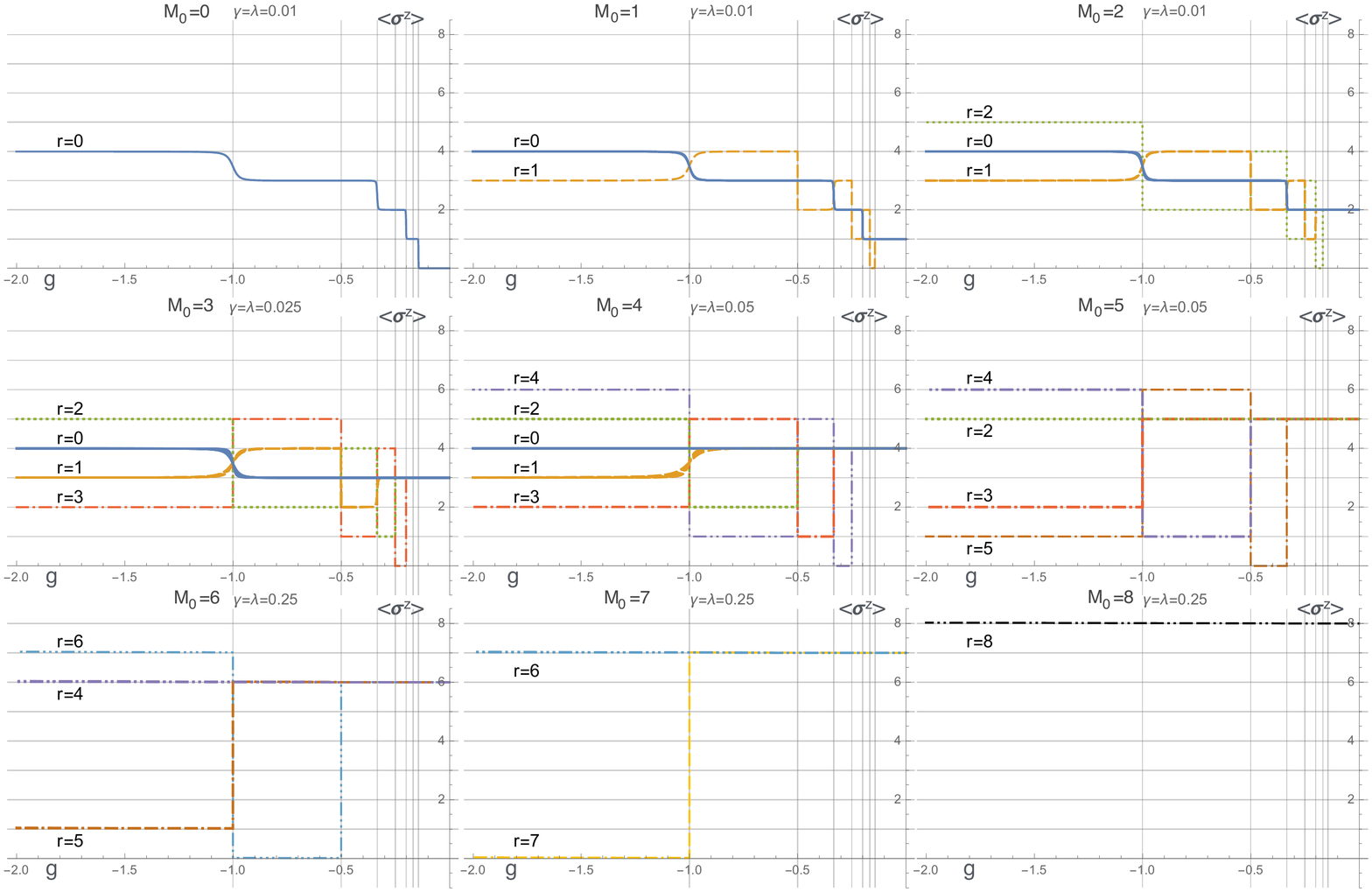}
\vspace{-0.2cm}
\caption{Z-axis magnetisation of the $2^N$ eigenstates as a function of the parameter $g$. The $\frac{N!}{(N-M)!M_0!}$ eigenstates whose parent state at $g=0$ has $M_0$ excitations are plotted in different panels. For a given $M_0$, all states with a given integer $r$ become (nearly) indistinguishable from one another. 
}
\label{allplots}
\end{center}
\end{figure}

Using this $g$-scanning algorithm for the superconductor's ground state, it was shown by Claeys et al. \cite{claeys_readgreen_2016} that, in the presence of weak coupling to a bath (superconductor) or an in-plane-magnetic field (central spin) ( $\lambda,\gamma \ne 0$), the ground state gets deformed in a such a way that it gains a single Cooper pair every time it goes through one the specific values of the coupling $g$ corresponding to the Read-Green points:

\bea
|g| = \frac{1}{N-2M-1}  \ \ \forall \ \ M = 0,1,... N/2,
\eea

\noindent at which $\frac{1}{|g|}$ correspond to an integer in the series $1,3,5 \dots N-1$. This specific step-like behaviour of the total number of excitations is shown in the upper left panel of FIG. \ref{allplots} of this work, while the corresponding expectation values of the individual spins can be seen in panel a) of FIG. \ref{sxsz}. If we had $\lambda,\gamma = 0$, the resulting number-conserving system would show a true energy level crossing between the $M$ and $M+1$ excitation-number sectors while, here, at and around those Read-Green point, the ground state hybridises between those two sectors of the Hilbert space. 

Since the quadratic equations (\ref{quadeq}) give us a simple access to the properties of individual eigenstates, the same study can be carried out, in a short amount of computation time, for every states in a small enough system. Here we choose to do so for the $2^N = 256$ states of a system of 8 spins since it is sufficient to reach clear conclusions about the system's generic structure. Fig. \ref{sxsz} presents the expectation values of $\sigma^z_i$ and $\sigma^x_i$ of 4 specific eigenstates as $g$ is varied. The parameters of the model were chosen as $\epsilon_i=i$, $\alpha=\beta=1$ and $\gamma = \lambda$ is used to make it such that, by symmetry, $\sigma^y_i$ behaves exactly as $\sigma^x_i$. 

The ground state presented in panel a) shows the behaviour described previously: gaining a single excitation each time  $\frac{1}{|g|}$ goes through odd integer values. As seen in \cite{claeys_readgreen_2016}, at these points, a strong resonance in the in-plane magnetisation $\left<\sigma_x\right>$ is also found indicating that individual spins are in a coherent superposition of their two $\sigma_z$ eigenstates: $\left|\uparrow_i\right>, \left|\downarrow_i\right>$. However, from the other eigenstates presented, we immediately see they too can undergo similar restructurations when going through those specific values of $\frac{1}{|g|}$. We first notice that  for the excited states of the system, these can occur at every integer valued $\frac{1}{|g|}$ between $1$ and $N-1$, whereas the ground state only gained an excitation at odd integer points. One also finds, by looking at the scale of the  plots, that the resonant $\left<\sigma_x\right>$ behaviour is found, in the presented cases, only for the states of panel a) and d), while panel b) and c) only show extremely weak in-plane expectation values. This can be easily understood since the full classification presented below will allow us to understand that the eigenstate in panels a) and d) hybridises between sectors containing $M$ and $M+1$ excitations while b) and c) involves two sectors which differ by more than one excitation, sectors between which $\sigma_x$ has no matrix element connecting them.

In order to characterise these (avoided) crossings and the excitation-number sectors that they involve, one can now turn to the expectation value of the total excitation number operator: $\frac{1}{2}\sum_{i=1}^N\left<\sigma^z_i+1\right>$. The behaviour of every one of the $2^N$ eigenstates is presented in FIG. \ref{allplots}.

As one can readily see, a limited number of possible behaviour are exhibited. Indeed, for large subsets of eigenstates, the plots are indistinguishable from one another, undergoing the exact same sequence of gains and losses of excitations as they go through the Read-Green points. For the 256 states plotted only 25 distinct behaviours are observed. Remarkably, each individual state's sequence of restructuration is entirely predictable by specifying only two integers defined by the structure of the $g=0$ parent state, namely the number of excitations it contains $M_0$ and a second integer $r$ (defined in the next section) which can also be computed in a simple way. 

\newpage

\onecolumngrid

\begin{table}[t]
\begin{center}
\includegraphics[width=17cm]{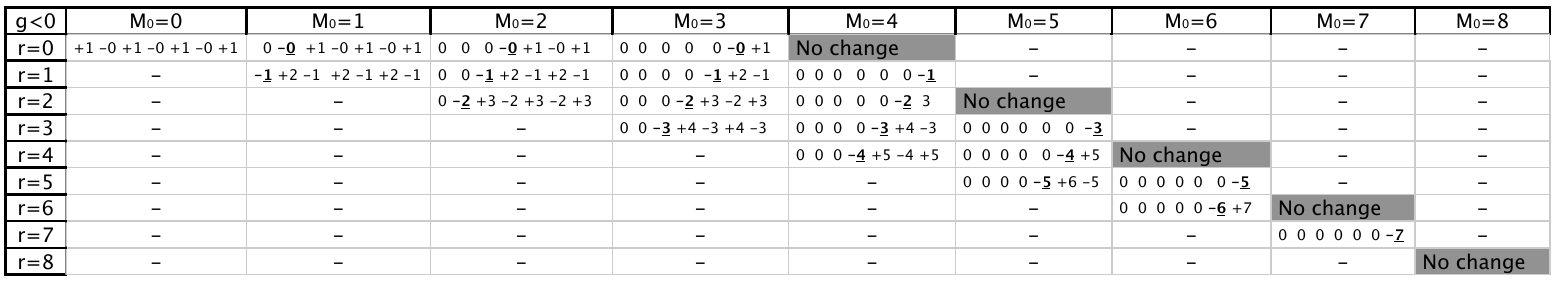}
\caption{The gains/losses are presented from left to right in order of increasing $|g|$, i.e. $g= \left(- \frac{1}{7}, -\frac{1}{6}, -\frac{1}{5}, -\frac{1}{4}, -\frac{1}{3}, -\frac{1}{2},-1,\right)$. The point at which the sequence (loss $-r$/ gain $r+1$ ) starts is underlined and bold. The greyed-out cells are those where no gain or loss of excitations occur since the start of the sequence would put it at a value of $g$ beyond the last Read-Green point $|g| =1$. The white cells correspond to values of $r$ which are impossible by construction.}
\end{center}
\end{table}

\twocolumngrid

\section{Classification of the profiles of magnetisation/number of pairs}

As seen in Fig. \ref{allplots}, each $g=0$ parent state defined by a given pair $(M_0,r)$ will have the same structure as $g$ is varied. Here, $M_0$ is simply the total number of up-spins (Cooper pairs) in the parent state, the integer $r$ can also be found directly by specifying the parent state's structure. It can be calculated by first separating the state into $P$ contiguous blocks which contain only ``down spins" on the left and ``up spins" on the right. For example, a parent state given by 
$$\circ \circ \bullet \bullet \bullet\circ \circ \bullet \circ \circ \bullet \bullet \bullet  \bullet \circ \circ  \circ$$ would be grouped into $P=3$ blocks:
$$\ \ \  \boxed{\circ \circ \bullet \bullet \bullet}\boxed{\circ \circ \bullet }\boxed{\circ \circ \bullet \bullet \bullet \ \bullet} \circ \circ \circ.$$

One then defines the excess number of ``up spins" in the rightmost block (numbered $P$) as $r_P = \mathrm{max}(N^{P}_{\bullet} - N^{P}_{\circ},0)$, with $N^{P}_{\bullet} $ the number of up and $N^{P}_{\circ}$ the number of down spins in block $P$ . One then moves on to block $P-1$ for which the number of spins up in excess is computed after carrying over the excess number from the preceding block so that $r_{P-1} = \mathrm{max}(N^{P-1}_{\bullet} + r_P - N^{P-1}_{\circ},0)$. The procedure is kept going by computing $r_{P-2} = \mathrm{max}(N^{P-2}_{\bullet} + r_{P-1} - N^{P-2}_{\circ},0)$ until the excess number from the last block gives us: $r \equiv r_1 = \mathrm{max}(N^{1}_{\bullet} + r_{2} - N^{1}_{\circ},0)$. In the example above, the rightmost block $P=3$ leads to $r_3 = 2$  (i.e.: 4 $\bullet$ - 2 $\circ$). The 2 up spins in excess are then added to the second block leading to $r_2 = 1$ (i.e.: 1 $\bullet$ + 2 $\bullet$ (from the third block) - 1  $\circ$ ). This excess spin is then added to the last block finally giving $r=2$ (i.e.:  3 $\bullet$  + 1 $\bullet$  (from the preceding block) - 2 $\circ$).  Interestingly, this specific integer $r$ has also been shown to give, for a given $g=0$ configuration, the number of Bethe roots which will diverge at large $g$ for any eigenstate of the isotropic XXX Richardson-Gaudin model \cite{faribaultdiv1,faribaultdiv2}.

As we now show, the specific sequence underwent by any $(M_0,r)$ state obeys relatively simple rules. As seen on Fig. \ref{allplots}, at the Read-Green points at which a $(M_0,r)$-state sees a loss of excitations it will always correspond to a loss of exactly $r$ excitations. Gains, on the other hand, always happen by gaining $r+1$ excitations. Moreover, they are always in strict alternance such that, moving from $g=0$, the state will, at a specific Read-Green point $g_s(M_0,r)$ first undergo a loss of $r$ excitation followed at the next Read-Green point by a gain of $r+1$. This sequence will be repeated until the last point at $g=1$ is reached. This statement is also true when $r=0$ which can then be understood as a "loss of zero excitation" followed by a gain of one as was the case for the superconducting ground state for example.

The last detail which remains to specify is the specific Read-Green point $g_s(M_0,r)$ at which this "loss of $r$/gain of $r+1$" sequence starts. It simple to verify that for every case where $r=M$ the loss/gain sequence begins specifically at the $M^\mathrm{th}$ Read-Green point (numbering them from 1 to $N-1$ in order of their magnitude $|g|$). For a given $M_0$ value, one then sees that when $r$ goes down by one (from $M_0$ to $M_0-1$, to $M_0-2$ and so on), the start of the sequence gets shifted to the next Read-Green point.  All in all, for $g<0$, any given eigenstate whose $g=0$ parent state is defined by $(M_0,r)$ will undergo an alternance of losses of $r$ excitations followed by gains of $r+1$ starting with a first loss at the $(2M_0-r)^{\mathrm{th}}$ RG point: 

\bea
g_s(M_0,r) =  \frac{1}{N-2M_0+r}.
\eea

Starting at half-filling $M_0=N/2$, the lowest possible value of $r$, namely $2M_0-N$, would place $|g_s|$ beyond the last Read-Green point and this small subset of states are the only ones which never undergo any such restructuration, i.e. they never have any (avoided) crossings with states from a different sector. For clarity, the presented case, $N=8$, is also detailed in TABLE 1.


Each of these losses/gains correspond, in the underlying number-conserving U(1)-symmetric models, to a level crossing between two orthogonal sectors with different number of excitations. These results therefore also provide a complete map of the magnetisation/filling factors sectors involved in the numerous degeneracies which occur at each of these Read-Green points in the excitation-number conserving systems. Using the Bethe Ansatz approach in the number conserving case, it was demonstrated that, at each Read-Green point there exists a duality which allows pairs of degenerate eigenstates to be created by adding, to the state in the lowest number sector, a given precise number of zero-energy excitations (Cooper pairs/Up-pointing spins) \cite{dunning, ibanez,rombouts}. The number of these zero-energy excitations defines a proper winding number which characterises the state's topology. In the problem treated here, by lifting the requirement of number conservation, each individual eigenstate, when deformed through a Read-Green point, is no longer required to have those zero-energy pairs and to go through the corresponding change in its topology. Indeed, states now simply loose (or gain) the corresponding number of excitations therefore avoiding the modification in their topological properties (topological phase transition).

The first point at $|g| = \frac{1}{N-1}$ involves only a single degeneracy between the $M_0=0$ state and the single $(M_0=1,r=1)$ state. However, as one progresses to Read-Green points at higher $|g|$, more and more states will become pairwise degenerate at the Read-Green point. At the last one ($|g|=1$) only a small minority of states (greyed-out cells) are not involved in a level crossing. While the analytical combinatorics would be fairly involved, one can check numerically for small values of even $N$ that  $\frac{N!}{\left(\left(N/2\right)! \right)^2}$ states would not undergo a level crossing at that point. While no Read-Green point will involve degeneracies over the whole spectrum which would define a true strong zero mode \cite{moran,fendley,alicea}, the fraction of states not involved in the ``strongest" zero mode (at $g=1$), namely $\frac{N!}{2^N \left(\left(N/2\right)! \right)^2}$, becomes vanishingly small in the $N \to \infty$ thermodynamic limit. 

For any given integrable Hamiltonian $H(g) = \sum_{i=1}^N \alpha_i R_i(g)$, accidental degeneracies between two states can actually occur at various values of $g$. Indeed, the energies $E_n=\sum_{i=1}^N \alpha_i r^n_i(g)$ and $E_m=\sum_{i=1}^N \alpha_i r^m_i(g)$ associated with two distinct eigenstates can easily become equal. However, the Read-Green points discussed in this work are radically different since they correspond the regular set of points $g$ at which two eigenstates can become degenerate for every possible integrable hamiltonian $H(g)$ of this class. 

\begin{figure}[h]
\begin{center}
\includegraphics[width=7.5cm]{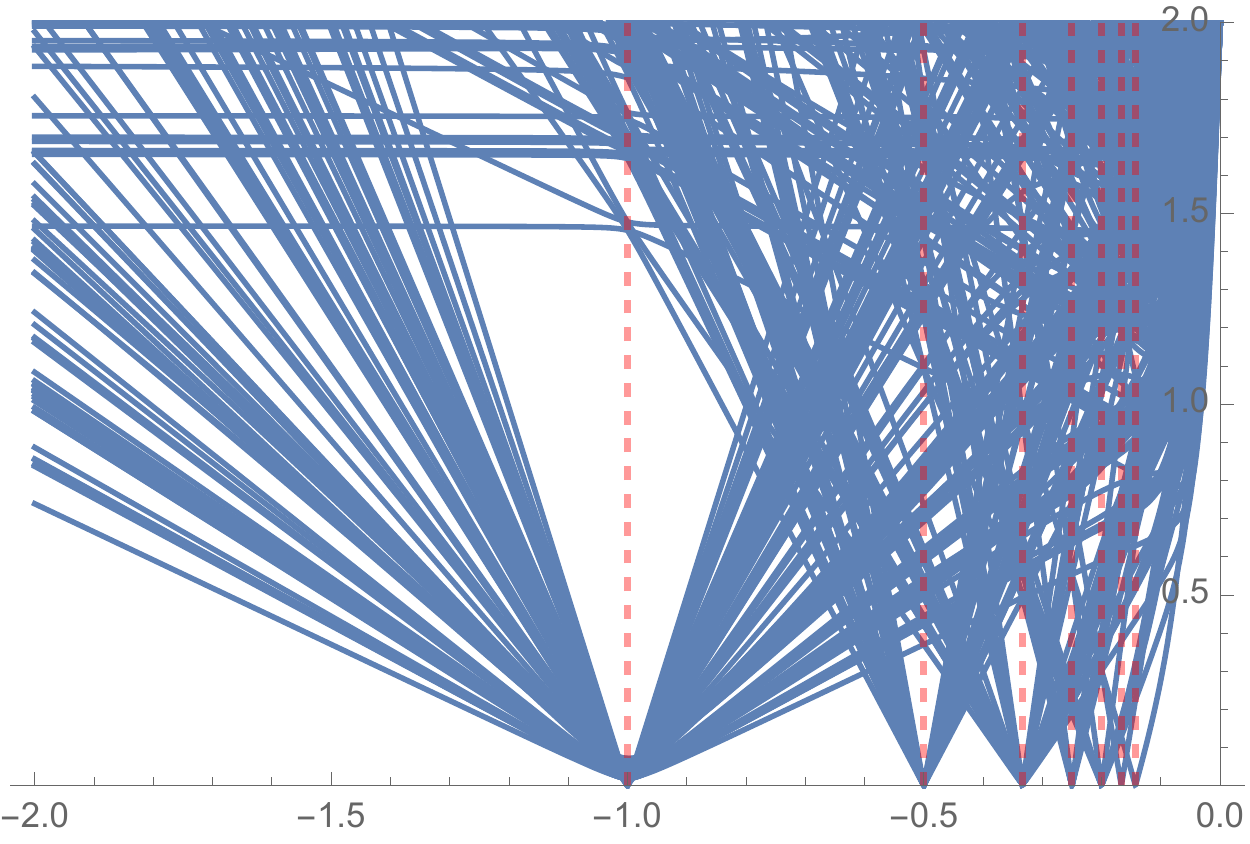}
\vspace{-0.2cm}
\caption{``Distance" $S_{nm}(g)$ between the set of eigenvalues for every pair of eigenstates. The zeros of these functions correspond to the points where eigenstates $n$ and $m$ share a set of eigenvalues which are common for every conserved charge. }
\label{degenplot}
\end{center}
\end{figure}

The states involved in the (avoided) crossings at these points become such that their full set of eigenvalues $(r^n_1, r^n_2 \dots r^n_N)$ and $(r^m_1, r^m_2 \dots r^m_N)$ are identical. For any pair of eigenstates (of index $n$ and $m$) the non-negative quantity $S_{nm}(g) = \displaystyle\sqrt{\sum_{i=1}^N \left(r_i^n(g)-r_i^m(g)\right)^2}$ can only become zero when both complete sets of eigenvalues coincide. By plotting this quantity for every pair of eigenstates, it is seen in FIG. \ref{degenplot} that these ``complete degeneracies" only happen at the Read-Green points. One can also clearly see in the figure that higher-coupling Read-Green points involve more and more of these degeneracies.

While this study has, so far, only focused on the the $g<0$ results, by symmetry, one can infer the corresponding $g>0$ behaviour. Indeed, at $g>0$, the conserved charges (\ref{conserved}) defining these systems are identical to those at $g<0$ after inversion of the z-axis: $\hat{z} \to -\hat{z}$. Consequently, after exchanging up-spins and  down spins ($\circ \leftrightarrow \bullet$), one can compute in the exact same way as $(M_0,r)$ the equivalent $(M_0^+,r^+)$ for any parent state, with $M_0^+ = N-M_m$. Since $z$ has been inverted, one then finds that the sequence will begin with a gain of $r_+$, followed by a loss of $r_+ + 1$ excitations with the sequence starting at the positive $g_s(M_0^+,r^+)$ Read-Green point. This is demonstrated in the next figure where three states sharing the same value of $r$ but different $r^+$ are plotted over the full range of positive and negative values of $g$.

\begin{figure}[h]
\begin{center}
\includegraphics[width=6.5cm]{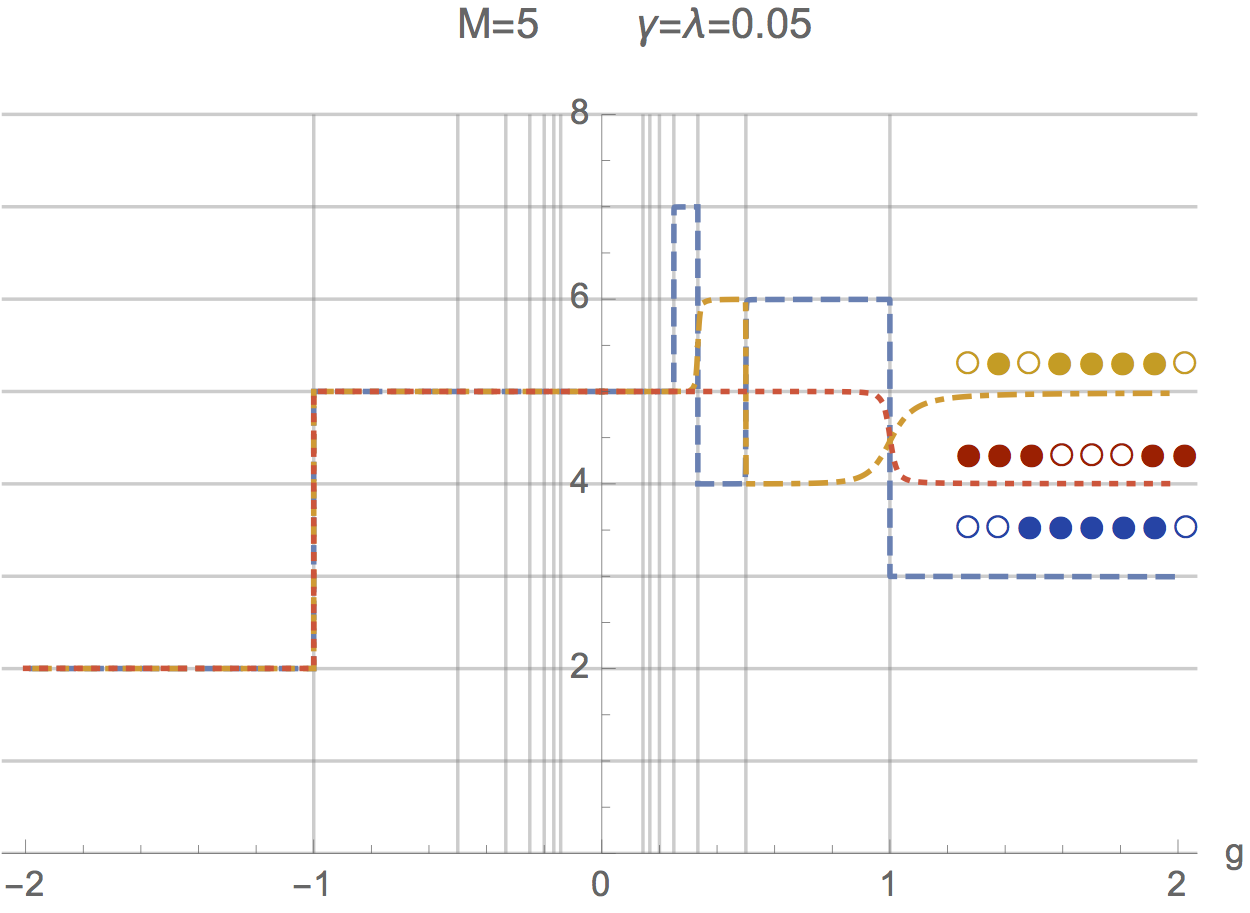}
\caption{ Total z-axis magnetization for specific states in the $M_0=5$ $(M_0^+ = 3)$:
$\circ\circ\bullet\bullet\bullet\bullet\bullet\circ$ ($r=3$ and $r^+=2$), $\bullet\bullet\bullet\circ\circ\circ\bullet \bullet$ ($r=3$ and $r^+=0$) and  $\circ\bullet\circ\bullet\bullet\bullet\bullet\ \circ$ ($r=3$ and $r^+=1$) . The vertical lines marks the Read-Green point at  $g=\pm1/n$ for $n=1,2,...7$
}
\end{center}
\end{figure}

\begin{figure} [h]
\includegraphics[width=5.5cm]{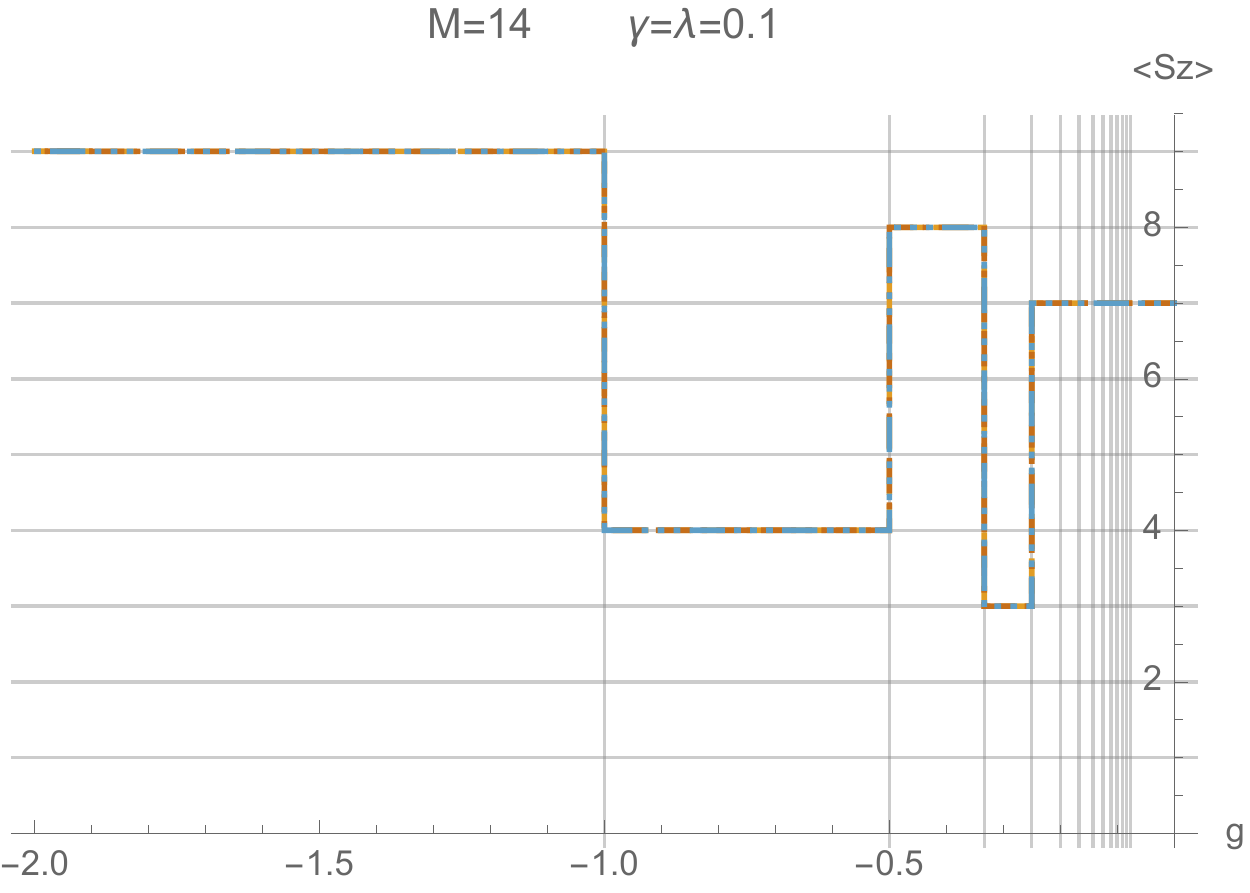}
\caption{Total magnetisation for N=14 spins, comparing the state $\bullet \bullet\bullet \circ\bullet \circ \bullet\bullet \circ\bullet \circ\circ\circ  \circ$ for the 3 distributions: $\epsilon_i = i$, $\epsilon_i = i^2$ and $\epsilon_i = \sqrt{i}$. The three curves are indistinguishable and correspond to the predicted result for $N=14, M_0=7, r = 4$, namely a sequence of  $-4$$/$$+5$ steps which starts at the $-(2M_0-r)=10^{\mathrm{th}}$ RG point: $g_s = - \frac{1}{N-2M_0+r} = - \frac{1}{4} $.
}
\label{larger}
\end{figure}

Finally we verify that the proposed result holds for larger system sizes and, since the Read-Green points have an underlying topological nature \cite{ibanez,rombouts,ortiz_topo1,ortiz_topo2,foster}, that the prescription holds true for arbitrary $\epsilon_i$, i.e.: namely different sets of XXZ integrable coupling constants. Such evidence is presented in FIG. \ref{larger}, where three systems of $N=14$ spins are compared for a given eigenstate. 

As does every other state, size or set of parameters we have numerically checked, the particular example presented here confirms the validity of our main result, not only in its capacity to predict the magnetisation sequence but also in its independence on the specific set of chosen coupling constants.

With the specific structure of avoided crossings now understood, it becomes possible to try to exploit it in order to create states with remarkably large number fluctuations in the system for example at $g=1$ where the $M=0$ and $M=N-1$ excitations sectors can hybridize. To do so, one would first prepare a $M=0$ state, which could be achieved by cooling down the system in a strong z-axis oriented external magnetic field ($\frac{1}{g} > N, \gamma=\lambda=0$), where the fully polarized $M=0$ state is the ground state. Instantaneously quenching down to weak magnetic field ($\frac{1}{g} < 1$), this initial state's would still project exclusively onto a single eigenstate of the new eigenbasis: the $M=0$ state (whose parent at $g=0$ is defined by $M_0=N-1$ and $r=N-1$). Adiabatically ramping the z-axis magnetic field back to $\frac{1}{g} =1$ would then, after turning on a perturbatively weak in-plane field, allow one to reach the $g=1$ state studied here. Since this state corresponds to the hybridisation of the $M=0$ and $M=N-1$ magnetisation sectors which should then show enormous magnetisation (Cooper pair number) fluctuations as it involves the fully down-polarised and the (nearly) fully up-polarised sectors. Since the $g=1$ point states at small excitation number $M$ systematically hybridises with states which contain a large number of excitations $N-M$, even an imperfect initial polarisation of the system would still exhibit such large magnetisation fluctuations since every sector at filling factor $\frac{M}{N} \equiv \rho < \frac{1}{2}$ has, at $g=1$ an avoided crossing with the sector at filling factor $1-\rho$.

\begin{figure} [h]
\includegraphics[width=5.5cm]{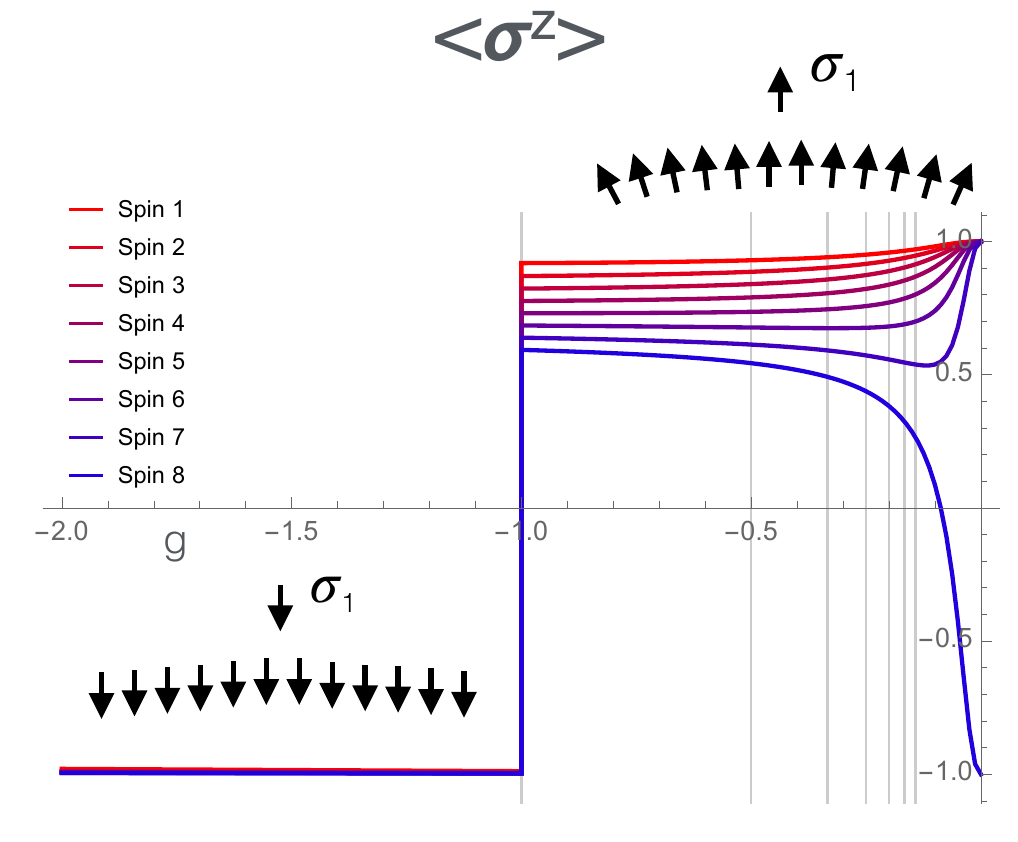}
\caption{The $M=N-1$, $r=N-1$ eigenstate (plotted here for $N=8$) whose remarkably large number fluctuations could, in principle, be observed through an instantaneous quench of the strong field ground state down to weak field, followed by an adiabatic ramping to the Read-Green point at $g=1$ where the state hybridises the sectors at filling factor $\rho=0$ and the one at filling $\rho=\frac{N-1}{N} \approx 1$.
}
\label{fluctuations}
\end{figure}

Moreover, as one can see in figure \ref{fluctuations}, in the $M=N-1$ sector the spin numbered 1 (which is the central spin for a hamiltonian given by $R_1$) is nearly completely up-polarised and so are the most strongly coupled environmental spins (spin 2, spin 3, ...). 

Only the most weakly coupled environmental spins will deviate from their spin-up state and for a large system size the $N-1$-excitation eigenstate would see its single down spin spread out over a larger bath making the strongly coupled spins even closer to perfect up-polarisation. Consequently, the two eigenstates involved in this hybridisation $\left|\Downarrow_1  \ \downarrow_2 \ \downarrow_3 \ \downarrow_4 \dots \downarrow_N\right> $ and $\left|\Uparrow_1  \ \uparrow_2 \ \uparrow_3 \ \uparrow_4 \dots \nearrow_N \right> $ could possibly be used as the two basis states of a spin qubit. Since the most strongly coupled environmental nuclear spins would then be systematically prepared in a way which mimics the central spin's state and would then act together as a large system coherently encoding the quantum information, such a setup could possibly provide strong protection against the decoherence induced by the environmental spin bath. Indeed, in this state, the available channels to flip down the central spin could only do so through the exchange terms which involves the most weakly coupled spin in the bath.

\section{Conclusion}

In this work we have shown how it is possible to fully characterise the z-axis magnetisation of every eigenstate of the XXZ Richardson-Gaudin models in the presence of a perturbatively weak X-Y plane magnetic field. These results also describe the number of Cooper pairs in a $p_x+ip_y$ topological superconductor weakly coupled to a particle bath. By ramping up the coupling constant $g$ or alternatively by ramping down the z-axis magnetic field, each state undergoes a series of gain/loss of magnetisation at the specific values known as Read-Green points. We demonstrate that each of those steps, their amplitude and the points at which they occur, when ramping up $g$ can be known in advance, for each given eigenstate, simply by knowning the spin configuration at $g=0$ which provides the two required integers $(M_0,r)$. These results provide a complete map of which sectors are involved in the numerous level crossings which occur in a magnetisation conserving XXZ model in a z-oriented field and, equivalently, in a closed Cooper-pair-number-conserving $p+ip$ topological superconductor.

\end{document}